\documentclass[epj]{svjour}
\usepackage{graphics}
\begin{document}
\title{Entropic force, holography and thermodynamics for static space-times}
\author{R. A. Konoplya\inst{1} 
}                     
%
%
\institute{Theoretical Astrophysics, Eberhard-Karls University of T\"{u}bingen, T\"{u}bingen 72076, Germany}
%
\date{Received: date}
%
\abstract{
Recently Verlinde has suggested a new approach to gravity which interprets gravitational interaction as a kind of entropic force.
The new approach uses the holographic principle by stating that the information is kept on the holographic screens which
coincide with equipotential surfaces. Motivated by this new interpretation of gravity (but not being limited by it) we study equipotential surfaces, the Unruh-Verlinde temperature, energy and acceleration for various static space-times: generic spherically symmetric solutions, axially symmetric black holes immersed in a magnetic field, traversable spherically symmetric wormholes of an arbitrary shape function, system of two and more extremely charged black holes in equilibrium. In particular, we have shown that the Unruh-Verlinde temperature of the holographic screen reaches absolute zero on the wormhole throat independently of the particular form of the wormhole solution.
\PACS{
      {PACS-key}{discribing text of that key}   \and
      {PACS-key}{discribing text of that key}
     } 
} 
\maketitle
\section{Introduction}
\label{intro}

A recent work by Verlinde \cite{Verlinde} has proposed a qualitatively new approach to gravity which joins the holographic and thermodynamic ideas.
The essence of the approach is based on two principles: first is the holographic principle through which space is emergent \cite{Holography}. The holographic principle supposes that space is only a kind of container of information. The information is stored on holographic screens, which separate points. The screens are equipotential surfaces of potentials built with the help of time-like Killing vectors.
The second principle suggests that when the body moves relatively the screen it exerts an entropic force, similar to the one that appears in osmosis or when big colloid molecules are surrounded by thermal environment of smaller particles. Thus, according to the Verlinde approach, gravity is not a fundamental interaction like other three but a kind of entropic force. Whatever fantastic these ideas look like, they allowed to derive both the Newton law of gravity and the Einstein equations \cite{Verlinde}!.

This gave recently an impetus for further study in this direction \cite{1}-\cite{16}. The cosmological implications have been considered in \cite{1}, \cite{3}, \cite{51}, \cite{7}, \cite{8}. In \cite{12}, \cite{13}, \cite{15}, \cite{151} the thermodynamic of some static and stationary black holes were considered.
In particular, in \cite{12} the temperature, acceleration and energy on the holographic screens were considered for Schwarzschild, Reissner-Nordstrom and Kerr solutions. Note also, that similar "reversing"  ideas about equipartition of energy in the horizon degrees of freedom
and the emergence of gravity was pronounced by T. Padmanabhan yet in 2004 \cite{Padmanabhan}.

Within Verlinde`s conception General Relativity seems to be remaining essentially unmodified on the macroscopic, observational level.
However, some geometrical objects, such as equipotential surfaces, Unruh temperature, and, foremost, the gravitational force acquire new interpretation.
A natural question which would arise is how various systems with strong gravity, such as black holes, wormholes,
neutron stars, or the whole universe look like in this new Verlinde`s conception.? An even more interesting question is how gravity interplays with fundamental matter fields.? An example of qualitative answer to the first question for a system of black holes could be seen on Fig. 4 of  \cite{Verlinde}, where it is demonstrated that maximum coarse graining happens at black hole horizons.

Having in mind this new description of gravity as an entropic force, and trying to answer at least partially the above two questions, in this paper we shall find holographic surfaces, acceleration, temperature and energy on them for various static solutions: general spherically symmetric static solutions, their particular case for traversable wormholes of arbitrary shape (Morris-Thorne wormholes) and of non-rotating black holes immersed in an asymptotically uniform magnetic field (described by the Ernst-Schwarzschild solution), the system of N extremely charged Reissner-Nordstrom black holes in equilibrium. In the last two cases the holographic surfaces are not spherically symmetric.

The paper is organized as follows: Sec II gives some basic formulas of the Verlinde approach. Sec III develops generic formulas for acceleration, Unruh temperature and energy of the holographic screens. In Sec IV the previous formulas are applied for the Morris-Thorne wormholes, while  Sec V is devoted to the Ernst-Schwarzschild black hole. Sec. VI constructs isotherms for the Majumdar-Papa-petrou solution for N centers.

\section{Basic relations}
%

In \cite{Verlinde} it is postulated that the change of entropy, related to the entropy that is saved on the holographic screen,
satisfies the following relations
\begin{equation}
\Delta S = 2 \pi k_{B}, \quad \Delta x = \hbar/m c.
\end{equation}
The coefficient $2 \pi$ is stipulated by matching the correct expression for the force $F$
\begin{equation}
F \Delta x = T \Delta S.
\end{equation}
The temperature is associated with the acceleration through the famous Unruh formula \cite{Unruh}
\begin{equation}
k_{B} T = \frac{\hbar a}{2 \pi c}.
\end{equation}
Implying the homogeneous distribution of information on the holographic screen,  for a particle approaching the screen one can write
\begin{equation}
m c^2 = \frac{1}{2} n k_{B} T,
\end{equation}
where $n$ is the number of bits. Together with the Unruh formula this gives
\begin{equation}
\frac{\Delta S}{n} = k_{B} \frac{a \Delta x}{2c^2}.
\end{equation}
In the general relativistic context one starts from a generalized form of the Newtonian potential
\begin{equation}
\phi = \frac{1}{2} \log (-g^{\alpha \beta }\xi_{\alpha} \xi_{\beta}),
\end{equation}
where $e^{\phi}$ is the red-shift factor that is supposed to be equal to unity at the infinity ($\phi =0$ at $r= \infty$), if the space-time is asymptotically flat. The background metric is supposed to be some static solution which admits a global time-like Killing vector $\xi_{\alpha}$.

The acceleration is defined by the formula
\begin{equation}
a^{\alpha} = - g^{\alpha \beta} \nabla_{\beta} \phi,
\end{equation}
and the Unruh-Verlinde temperature on the screen is given by the formula
\begin{equation}\label{T}
T = \frac{\hbar}{2 \pi} e^{\phi} n^{\alpha} \nabla_{\alpha} \phi,
\end{equation}
where $n_{\alpha}$ is a unit vector, that is normal to the holographic screen and the Killing time-like vector $\xi_{\beta}$.

\section{Static spherically symmetric solution}

Let us consider the static spherically symmetric space-times of the generic form
\begin{equation}
d s^2 = - A(r) d t^2 + B(r) d r^2 + C(r) r^2 (d \theta^2 + \sin^2 \theta d \phi^2).
\end{equation}
The ($\alpha=0$, $\beta=1$)-component of the Killing equations
\begin{equation}
\partial_{\alpha} \xi_{\beta} +  \partial_{\beta} \xi_{\alpha} = 2 \Gamma_{\alpha \beta}^{\gamma} \xi_{\gamma}
\end{equation}
gives us one of the Killing vectors
\begin{equation}
\xi_{\alpha} = (C e^{2 \int \Gamma_{01}^{0}dr}, 0, 0, 0), \quad \Gamma_{01}^{0} = A' /2 A.
\end{equation}
Looking for a time-like Killing vector and using the other components of the Killing equations we can choose $C = -1$.
\begin{equation}
\xi_{\alpha} = (-A(r), 0, 0, 0).
\end{equation}
The potential and the acceleration (up to the choice of sign) have the form
\begin{equation}\label{pp}
\phi = \frac{1}{2} \log(A(r))
\end{equation}
\begin{equation}\label{aa}
a = (0, A'(r)/2 A(r)B(r), 0, 0).
\end{equation}
The last formula has been recently derived in \cite{12} with a typo in the numerical factor.

Using the equation (\ref{T}), the Unruh-Verlinde temperature can be written in the form
\begin{equation}\label{tt}
T = \frac{\hbar}{2 \pi} e^{\phi} \sqrt{g^{\alpha \beta} \phi_{,\alpha} \phi_{,\beta}} = \frac{\hbar}{4 \pi} \frac{A'(r)}{\sqrt{A(r) B(r)}}
\end{equation}

For spherically symmetric space-times this immediately produces the formula for the energy on the holographic screen $S$
\begin{equation}\label{ee}
E = \frac{1}{4 \pi} \int_{S} e^{\phi} \nabla \phi d A = 2 \pi r^2 \hbar^{-1} T.
\end{equation}

Now we are in position to consider a particular example of the static spherically symmetric solutions: Morris-Thorne wormholes.

\section{Lorentzian traversable wormholes}

Spherically symmetric lorentzian traversable wormholes of arbitrary shape can be modeled by a Morris-Thorne anzats \cite{Morris-Thorne}
\begin{equation}\label{MT}
ds^2 = - e^{2 \Phi (r)} dt^2 + \frac{d r^2}{1 - \frac{b(r)}{r}} + r^2 (d \theta^2 + \sin^2 \theta d \phi^2).
\end{equation}
Here $\Phi(r)$ is the lapse function which determines the red-shift effect and tidal force of the wormhole space-time.
The $\Phi = 0$ wormholes do not produce the tidal force. A shape of a wormhole is completely determined by another function $b(r)$, called the shape function.
A wormhole`s throat is situated at a minimal value of $r$, $r_{min} = b_0$. Thus the coordinate $r$ runs from $r_{min}$ until spatial infinity $r = \infty$, while in the proper radial distance coordinate $d l$, given by the equation
\begin{equation}
\frac{dl}{d r} = \pm \left(1- \frac{b(r)}{r}\right)^{-1/2},
\end{equation}
there are two infinities $l= \pm \infty$ at $r= \infty$.
From the requirement of absence of singularities, $\Phi(r)$
must be finite everywhere, and the requirement of asymptotic flatness gives $\Phi(r) \rightarrow 0$ as $r \rightarrow \infty $
(or $ l \rightarrow \pm \infty$). The shape function $b(r)$ must be such that $1- b(r)/r > 0$ and
$b(r)/r \rightarrow 0$ as $r \rightarrow \infty $ (or $ l \rightarrow \pm \infty$). In the throat $r = b_0$ and, therefore, $1-b(r)/r$ vanishes. Probing of such wormhole geometry by test fields propagating in its background has been recently done in \cite{Konoplya-Zhidenko-worm1}.

The time-like Killing vector has the form
\begin{equation}
\xi_{\alpha} = (- e^{2 \Phi(r)}, 0, 0, 0),
\end{equation}
that satisfies $\xi_{\alpha} \xi^{\alpha} = -1$ at infinity. Here we imply such form of $\Phi(r)$ that provides existence of the time-like Killing vector everywhere in the space.

Using the formulas (\ref{pp}), (\ref{aa}), (\ref{tt}) one can easily find the expressions for the potential, acceleration, and Unruh temperature,
\begin{equation}
\phi = \Phi(r),
\end{equation}
\begin{equation}
a = (0, -(1- b(r)/r) \Phi'(r), 0, 0)
\end{equation}
\begin{equation}
T = \frac{\hbar}{2 \pi} \Phi'(r) e^{\Phi} \sqrt{1- \frac{b(r)}{r}}
\end{equation}
From the formulas above we conclude that both the acceleration and the Unruh temperature are vanishing on the throat of the wormhole.
As one can easily see this feature is quite general and does not depend on the particular form of a wormhole.
Using Eq. (\ref{ee}), one find  that the energy on the holographic screen (which is spherically symmetric in this case) also vanishes on the throat
\begin{equation}
E = \Phi'(r) e^{\Phi} \sqrt{1- \frac{b(r)}{r}} r^2.
\end{equation}
The vanishing of the energy and temperature on the wormhole throat could probably be explained in some way within the holographic principle.

\section{Black holes immersed in a magnetic field}

The metric of a non-rotating black hole immersed in an asymptotically uniform magnetic field can be described
by the Ernst-Schwarzschild solution \cite{Ernst}
$$ d s^{2} = -\Lambda^{2} \left(
f(r) d t^{2} - f(r)^{-1} d
r^{2} -r^{2} d \theta^{2}
\right)$$
\begin{equation}
 + \Lambda^{-2} r^{2} \sin^{2} \theta  d \phi^{2}, \quad f(r) = \left(1- \frac{2 M}{r} \right),
\end{equation}
where the external magnetic field is determined by the real
parameter $B$, and
\begin{equation}
\Lambda = 1 + \frac{1}{4} B^{2} r^{2} \sin^{2} \theta.
\end{equation}
The vector potential for the magnetic field is given by the
formula:
\begin{equation}
A_{\mu} d x ^{\mu} = \frac{B r^{2} \sin^{2} \theta} {2 \Lambda}
d \phi.
\end{equation}

As a magnetic field is assumed to exist everywhere in space, the
above metric is not asymptotically flat.
However one can consider this solution only in some region near black hole and match it with some asymptotically
flat solution in far region. This would provide, if one wishes, asymptotic flatness of infinity.

The event horizon is again $r_{h} = 2 M$, and the surface gravity at the event horizon
is the same as that for the Schwarzschild metric,
\begin{equation}
\chi = 2 \pi T_{H} = \frac{1}{4 M}.
\end{equation}
This leads to the same expression for the Hawking temperature $T_{h}$ as for the
case of the Schwarzschild black hole \cite{Radu}.
Further properties of this solution were investigated in \cite{MagnetizedBH}.

The time-like Killing vector has the following form
\begin{equation}
\xi_{\alpha} = (-(1-2 M/r) \Lambda^2, 0, 0, 0),
\end{equation}
what leads immediately to the potential
\begin{equation}
\phi = \frac{1}{2} \log ((1-2 M/r) \Lambda^2).
\end{equation}
The non-zero components of the acceleration are
\begin{equation}\label{amag}
a^1 = -\left(1-\frac{2 M}{r}\right)\frac{\Lambda^{-2}}{r}\left(-2 + 2 \Lambda^{-1} - \frac{M}{r- 2 M}\right),
\end{equation}
\begin{equation}
a^2 = \frac{1}{4 r^2 \Lambda^3} B^2 r^2 \sin^2 \theta.
\end{equation}
The Unruh-Verlinde temperature equals
\begin{equation}\label{Tmag}
T = \frac{\hbar}{2 \pi} \frac{\sqrt{P(r)}}{r^2 (4 + B^2 r^2 \sin^2 \theta)},
\end{equation}
where 
$$ P(r) = 16 M^2 + B^4 (3 M- 2 r)^2 r^4 \sin^4 \theta +$$
\begin{equation}
4 B^2 r^2 (- 6 M^2 + 4 M r + B^2 r^3 (r-2 M) \cos^2 \theta) \sin^2 \theta.
\end{equation}

\begin{figure*}
\resizebox{0.8\textwidth}{!}{%
\includegraphics{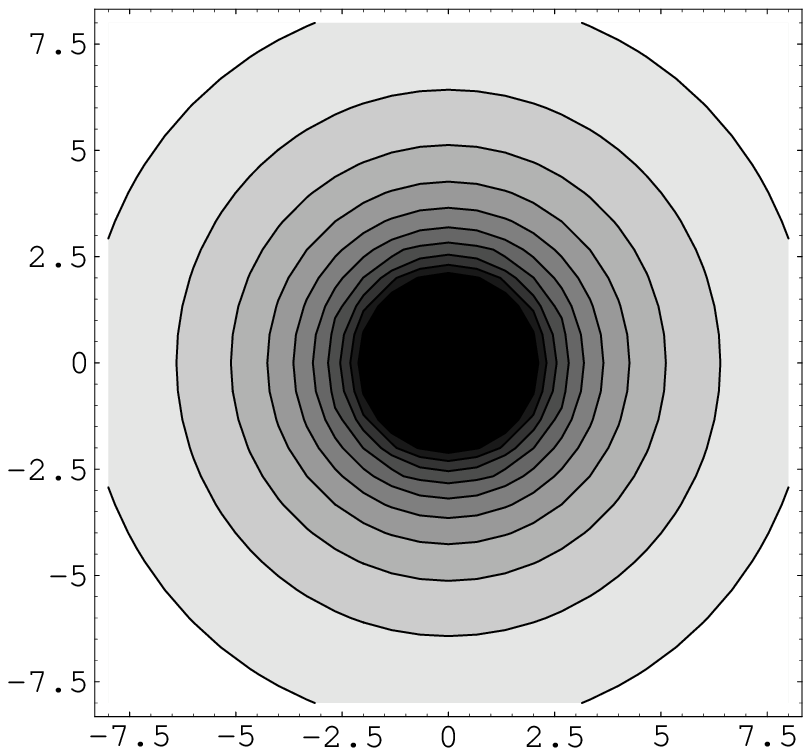},\includegraphics{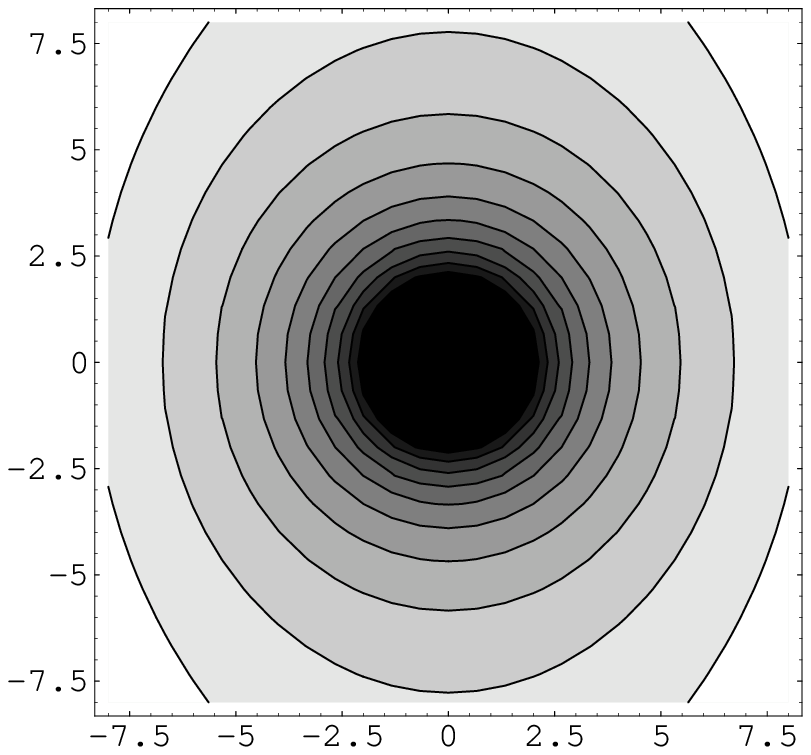},\includegraphics{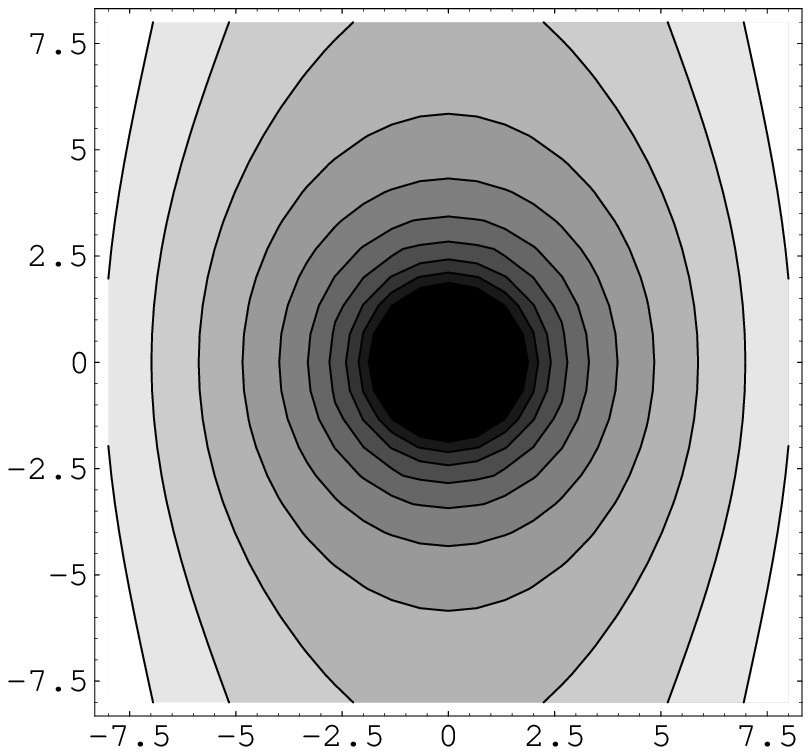}
}
\caption{The potential $e^{2 \phi}$ as a contour plot in (x, y) coordinates: $B=1/100$, $1/20$, $1/10$; $M=1$. The holographic screens are situated on the equipotential surfaces $\phi(x, y) = const$. The magnetic field is directed vertically.}
\label{111}       
\end{figure*}
\begin{figure*}
\resizebox{0.8\textwidth}{!}{%
\includegraphics{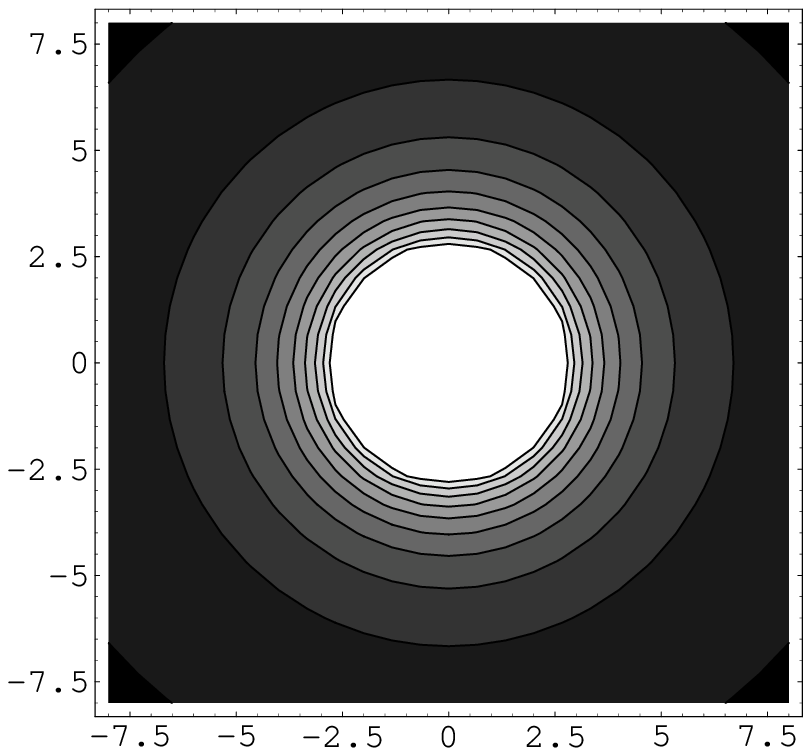},\includegraphics{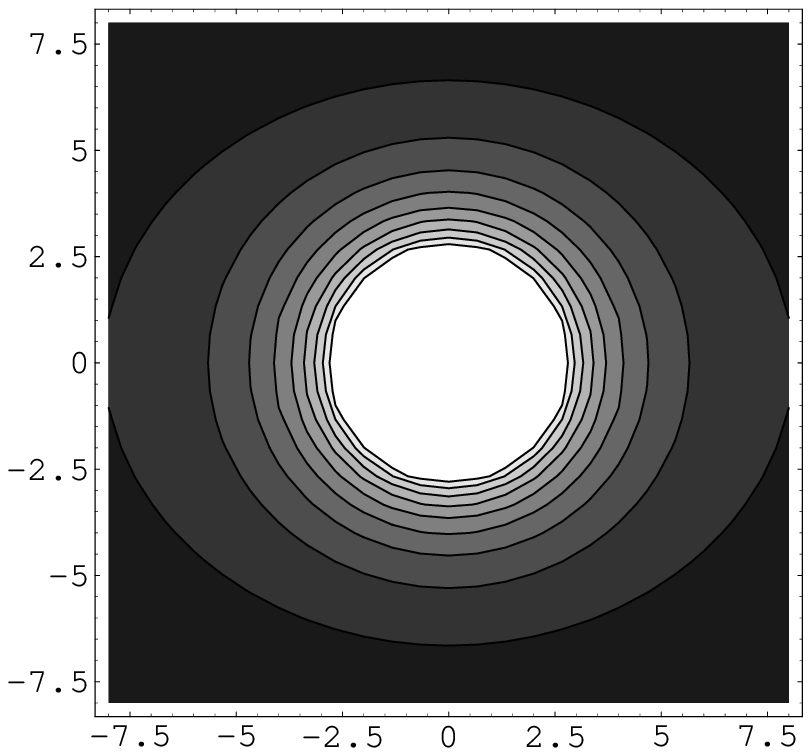},\includegraphics{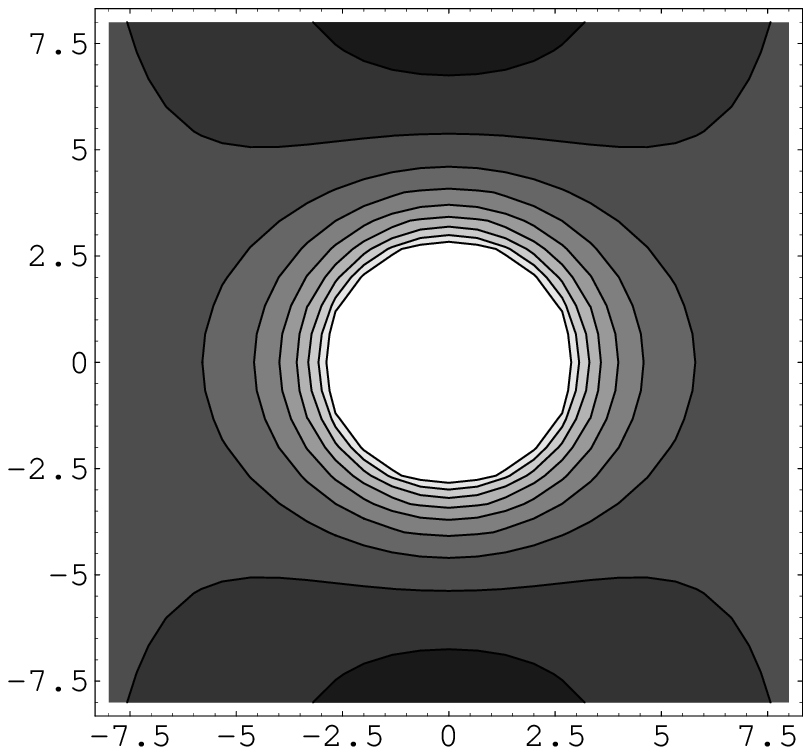}
}
\resizebox{0.8\textwidth}{!}{%
\includegraphics{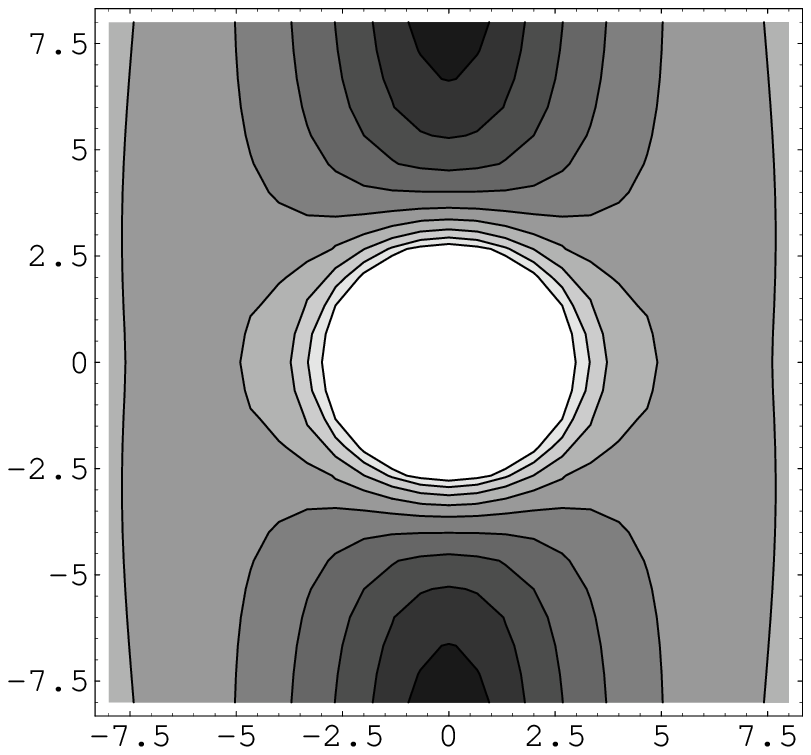},\includegraphics{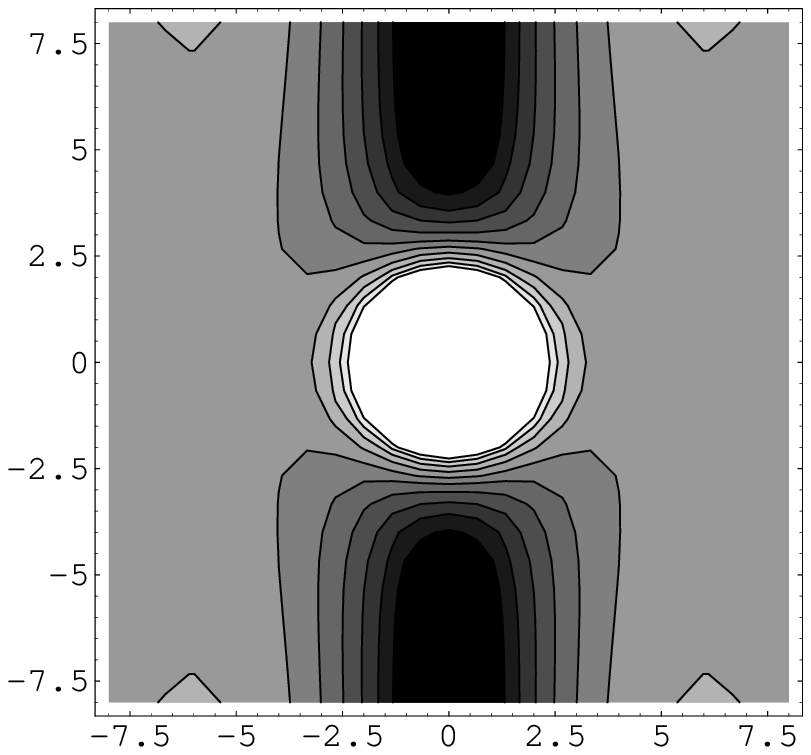},\includegraphics{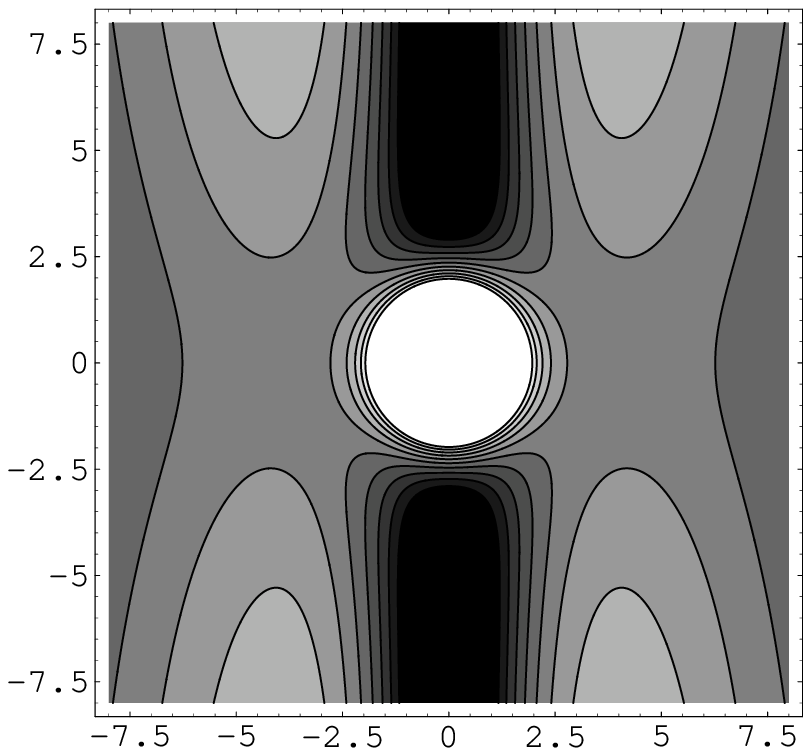}
}
\caption{The Unruh temperature $T$ multiplied by $2 \pi/\hbar$ in the (x, y) coordinates: $B=1/100$, $1/20$, $1/10$, $1/5$, $1/3$, $1/2$, $M=1$.
The magnetic field is directed vertically.}
\label{111}       
\end{figure*}
\begin{figure*}
\resizebox{0.8\textwidth}{!}{%
\includegraphics{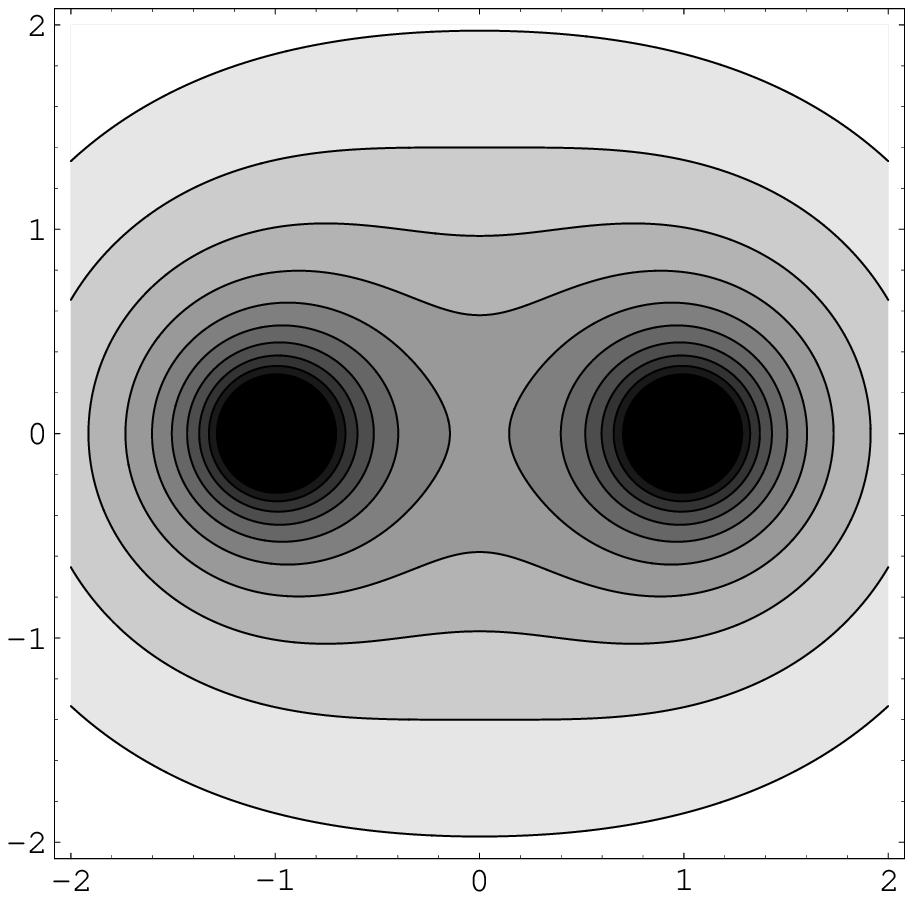},\includegraphics{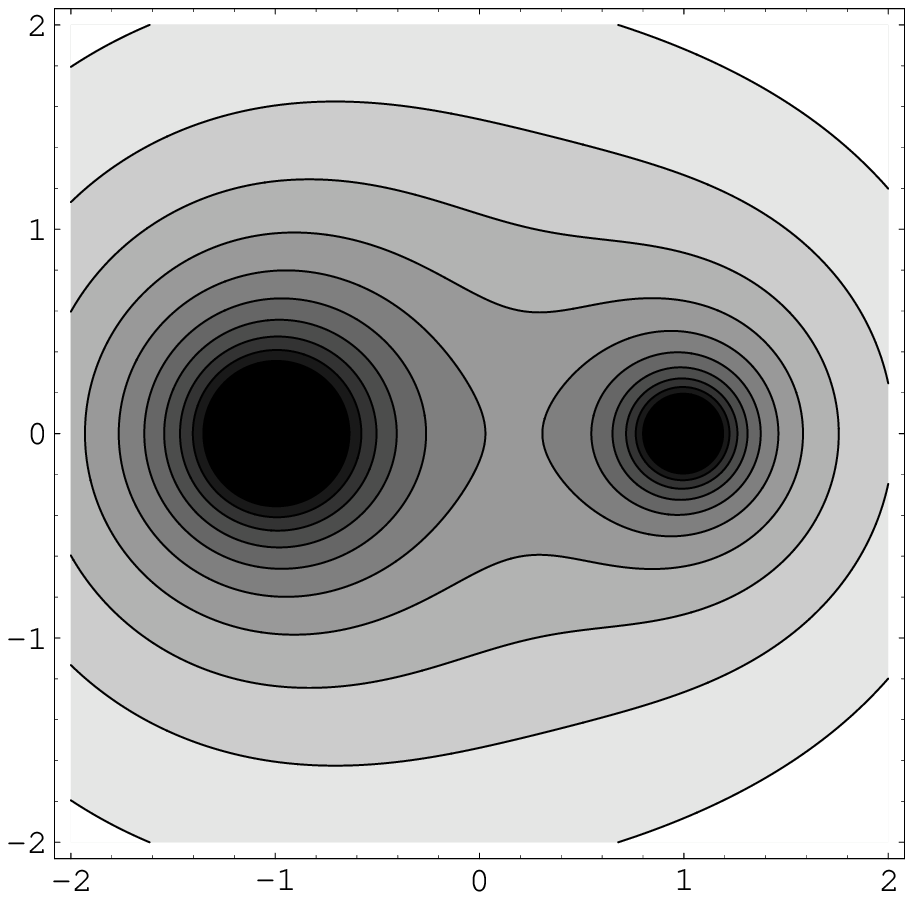},\includegraphics{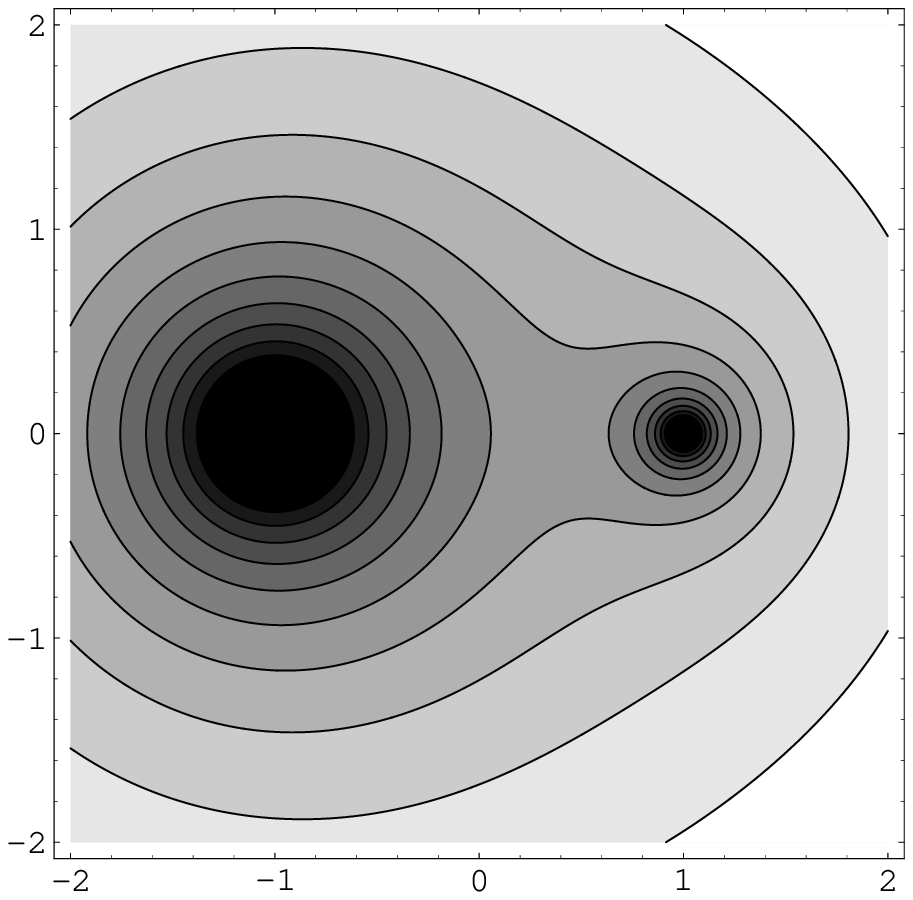}
}
\caption{Equipotential lines of $\phi$ in the $(x, y)$-plane for system of two extremely charged black holes: $M_{2} =1$, $x_1 =-1$, $x_2 =1$; a) $M_{1} = 1/3$ -left b) $M_{1} = 2/3$ c) $M_{1} = 5/3$ - right ($\hbar=1$).}
\label{00}       
\end{figure*}

In the limit $M=0$ we find the values of temperature and $a^1$-component of the acceleration for pure magnetic universe described by the Melvine metric,
\begin{equation}
a^{1} = \frac{2}{\Lambda^2 r}\left(1- \frac{1}{\Lambda}\right),
\end{equation}
\begin{equation}
T =\frac{\hbar}{\pi} \frac{B^2 r \sin \theta}{4 + B^2 r^2 \sin^2 \theta}.
\end{equation}
The $\theta$-component of acceleration remains the same and is independent on the presence of a black hole.
In the limit $B=0$ the above formulas (\ref{Tmag}),(\ref{amag}) reduce to their Schwarzschild values \cite{12}.

Let us note that the Unruh temperature on the event horizon $r = 2 M$ coincides with the Hawking temperature of the black hole $T = 1/ 8 \pi M$.
The equipotential surfaces representing the holographic screens are shown on Fig. 1 in the Cartesian coordinates (x, y) instead of polar ($r$, $\theta$). There one can see that the maximum coarse graining of the information occurs on the black hole horizons. The Unruh temperature as a function of coordinates can be seen on the contour plot Fig.2. When the magnetic field is very small $B \ll M$, the holographic surfaces and lines of constant temperature have almost spherical form. The strong deformations start when $B \simeq M$. Then, one can see that the region of low temperature is formed along the magnetic field axis outside the black hole.

\section{Two, three and N black holes: Majumdar-Papapetrou solution}

A system of two black holes can be supported in static equilibrium if black holes are extremely charged in such a way that the gravitational attraction is balanced by the electromagnetic repulsion. The static exact solution was found by Majumdar \cite{Majumdar} and Papapetrou \cite{Pappaetrou}. Hartle and Hawking showed that this solution describes two extremely charged black holes in equilibrium \cite{Hartle}. Later the generalization for $N$ black holes had been found \cite{GibbonsN}.
The metric has the form,
\begin{equation}\label{MP1}
ds^2 = -U^{-2}(x, y, z) dt^2 + U^{2}(x, y, z) (d x^2 + d y^2 + d z^2),
\end{equation}
where
\begin{equation}\label{MP2}
U = 1 + \sum_{i =1}^{N} \frac{M_{i}}{\sqrt{(x-x_i)^2 + (y-y_i)^2 + (z-z_i)^2}}.
\end{equation}
Here $M_i$ are masses of extreme Reissner-Nordstrom black holes, $N$ is the number of black holes, ($x_i$, $y_i$, $z_i$) is location of the i-th black hole.

The metric  (\ref{MP1}),(\ref{MP2}) is static and allows for the time-like Killing vector $$\xi_{\alpha} = (-U^{-2}, 0, 0, 0)$$. The gravitational potential is
\begin{equation}
\phi(x, y, z) = \log(U^{-1}).
\end{equation}
The non-zero components of the acceleration are
\begin{equation}
a^{i} = \frac{1}{U^3} \frac{\partial U}{\partial x^{i}}, \quad i = 1, 2, 3.
\end{equation}
The temperature can be found by the formula (\ref{T}),
\begin{equation}
T = \frac{\hbar}{2 \pi U^{3}} \sqrt{\sum_{j=1}^{3}\left(\frac{\partial U}{\partial x^{j}}\right)^2}.
\end{equation}

\begin{figure*}
\resizebox{0.8\textwidth}{!}{%
\includegraphics{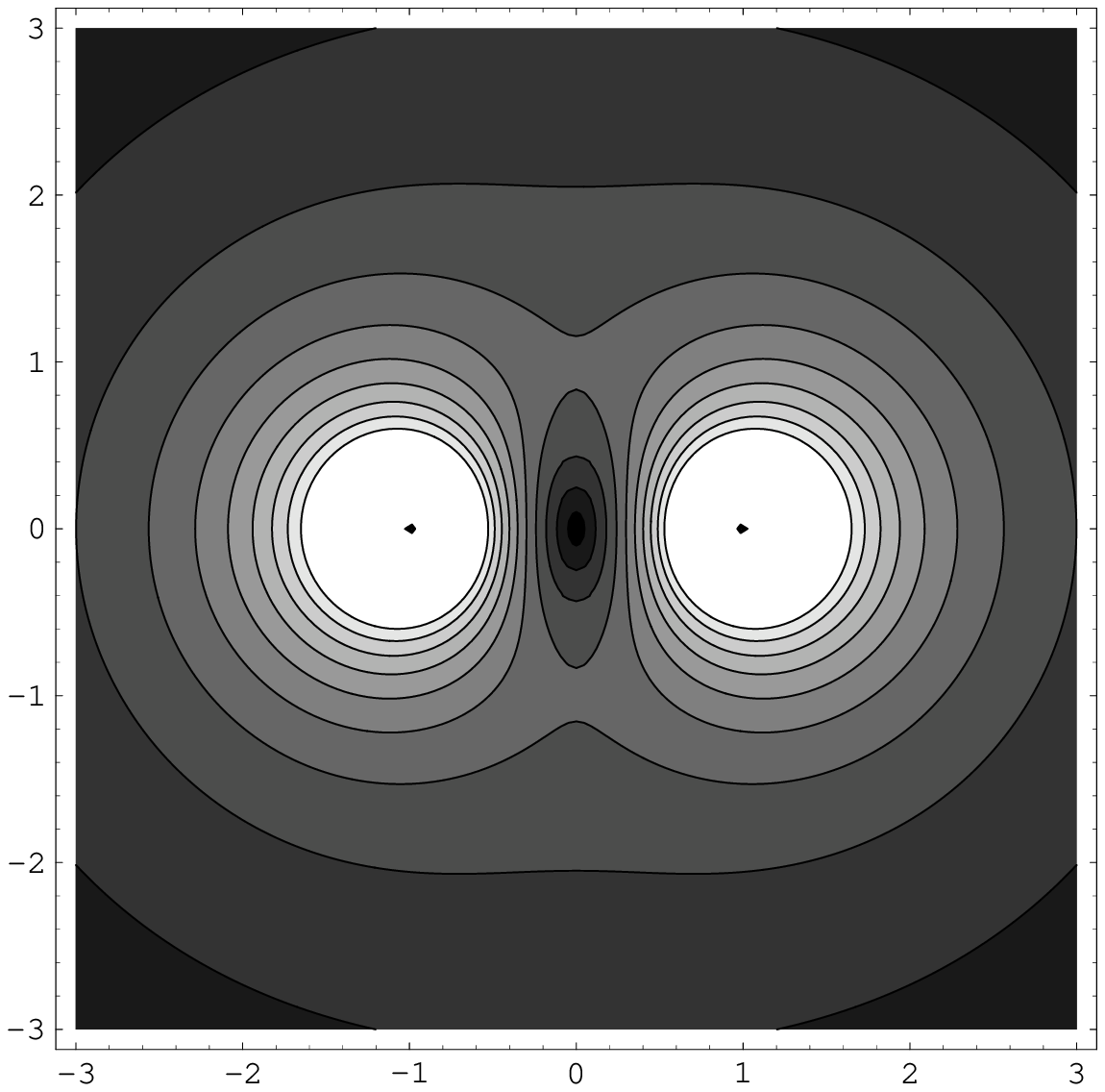},\includegraphics{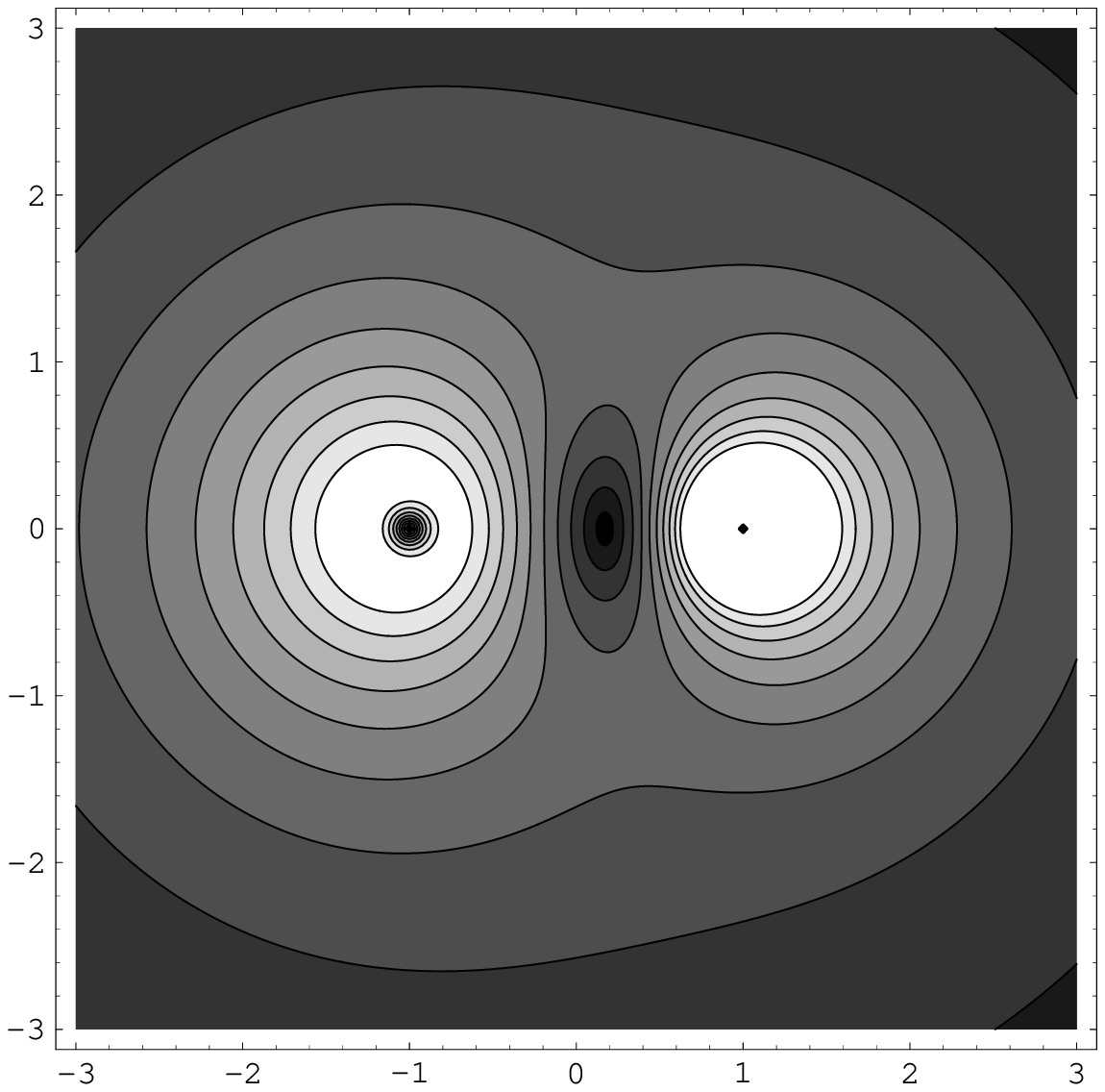},\includegraphics{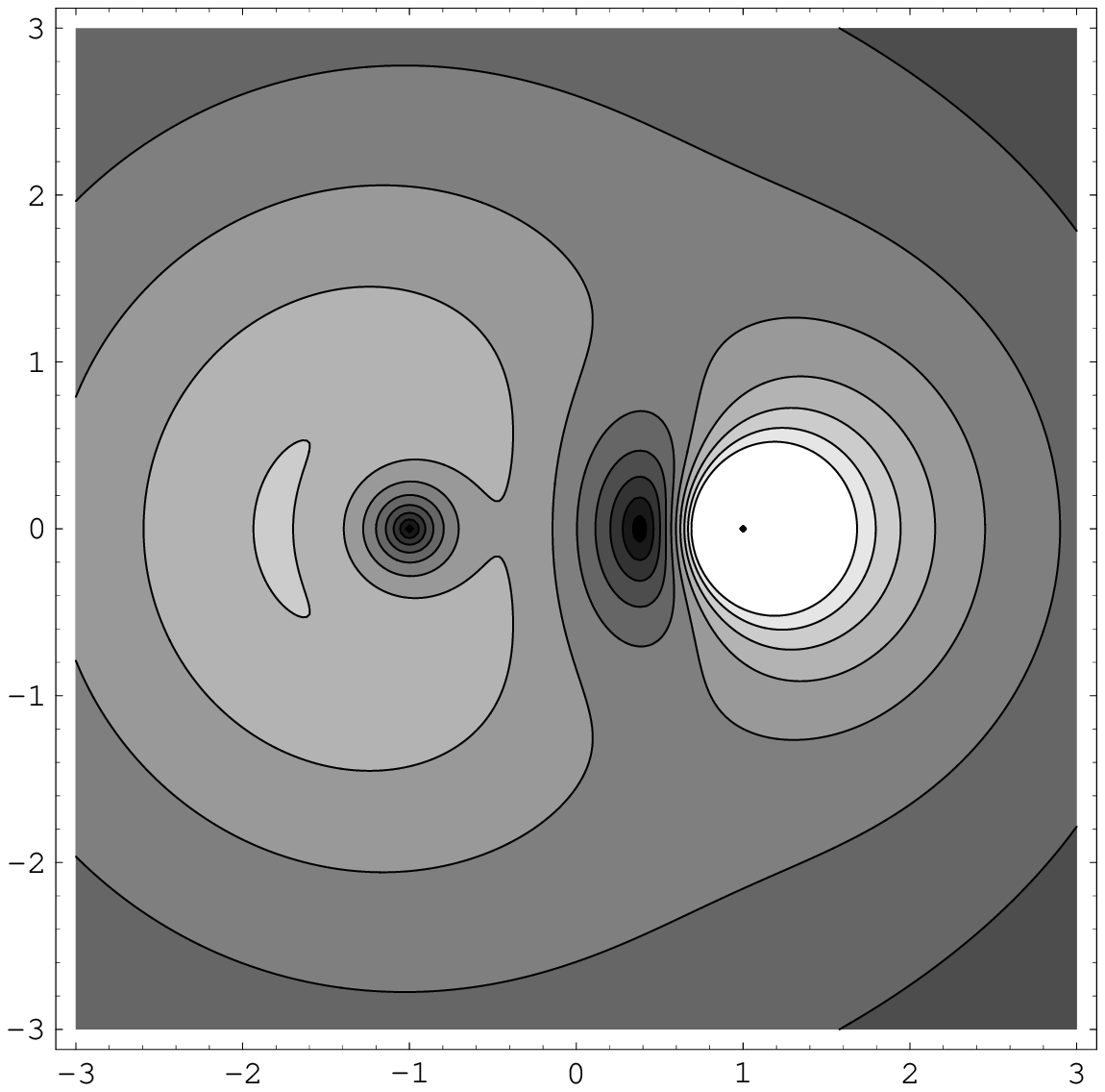}
}
\caption{Temperature for system of two extremely charged black holes in the $(x, y)$-plane: $M_{2} =1$, $x_1 =-1$, $x_2 =1$; a) $M_{1} = 1/3$ -left b) $M_{1} = 2/3$ c) $M_{1} = 5/3$ - right ($\hbar=1$)}
\label{AA}
\end{figure*}

\begin{figure*}
\resizebox{0.8\textwidth}{!}{%
\includegraphics{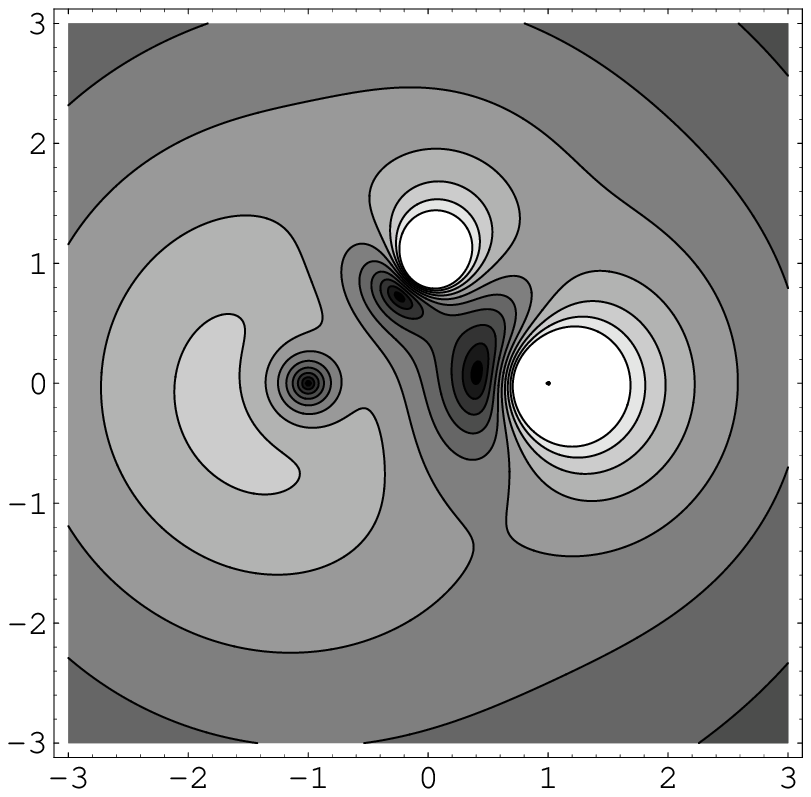},\includegraphics{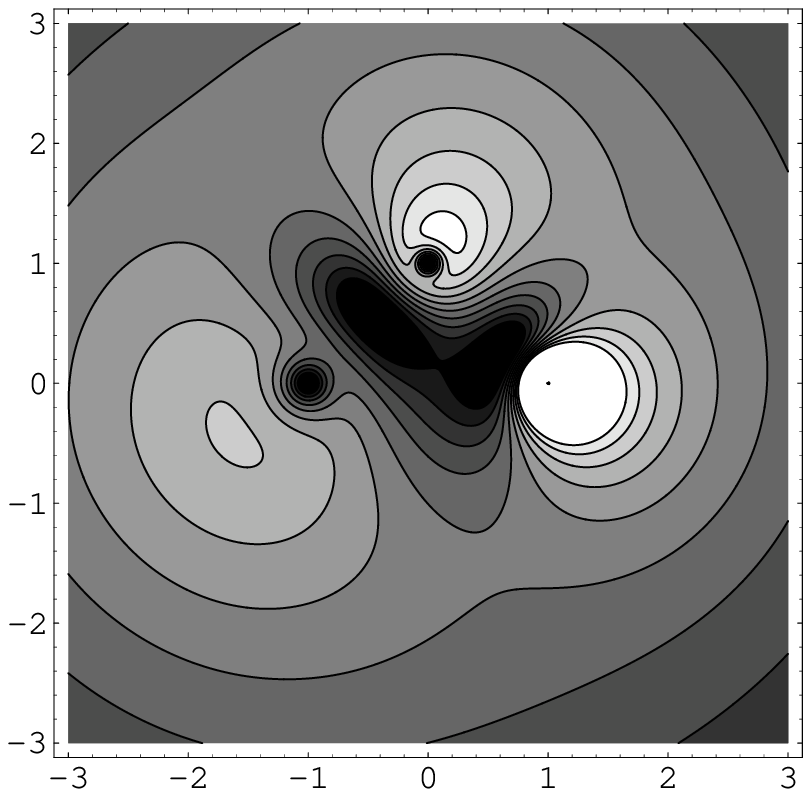},\includegraphics{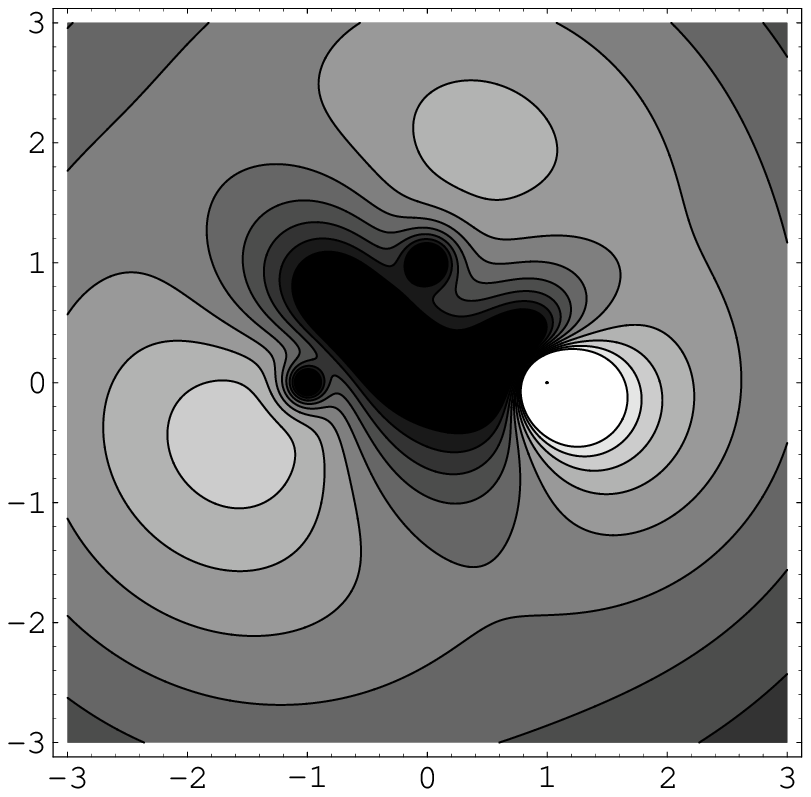}
}
\caption{Temperature for system of three extremely charged black holes in the $(x, y)$-plane: $M_{1} =5/3$, $M_2 =1/3$, $x_1 =-1$, $x_2 =1$, $y_3 =1$; $M_{3} =1/5$ (left), $M_{3} =1$, $M_{3} =2$ (right) ($\hbar=1$).}
\label{BB}
\end{figure*}

Contour plots of temperature for systems of two (Fig. \ref{AA}) and three (Fig. \ref{BB}) black holes show quite peculiar picture of spacial distribution of temperature in space. There one can see that due-to counter actions of black holes there is a region of relatively low temperature somewhere \emph{in between} the black holes. This is quite natural, taking account of equivalence between temperature and
acceleration (i.e. force), as the region of low temperature corresponds to the region of small acceleration, where gravitational attraction of a test particle by each black hole is well balanced.

The above results can be immediately generalized for the dilatonic N-black hole solution,
$$ ds^2 =  -U^{-2/(1+a^2)}(x, y, z) dt^2 + $$
\begin{equation}
U^{2/(1+a^2)}(x, y, z) (d x^2 + d y^2 + d z^2),
\end{equation}
where
\begin{equation}
U = 1 + \sum_{i =1}^{N} \frac{M_{i}(1+a^2)}{\sqrt{(x-x_i)^2 + (y-y_i)^2 + (z-z_i)^2}}.
\end{equation}
The temperature for the dilatonic case equals
\begin{equation}
T = \frac{\hbar}{2 \pi (1+a^2) U^{(3+a^2)/(1+a^2)}} \sqrt{\sum_{j=1}^{3}\left(\frac{\partial U}{\partial x^{j}}\right)^2}.
\end{equation}


\begin{figure*}
\resizebox{0.8\textwidth}{!}{%
\includegraphics{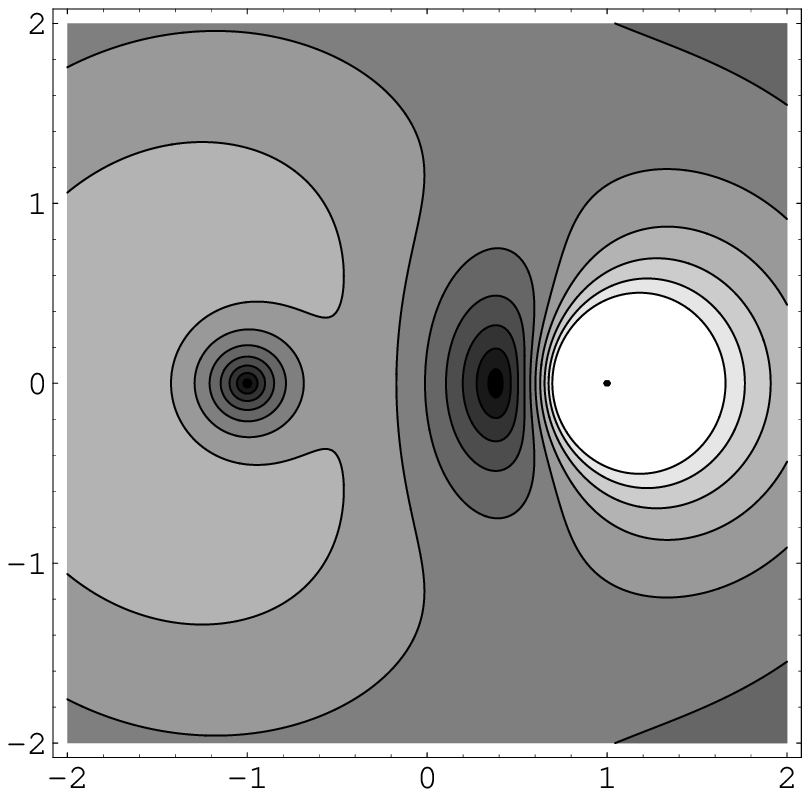},\includegraphics{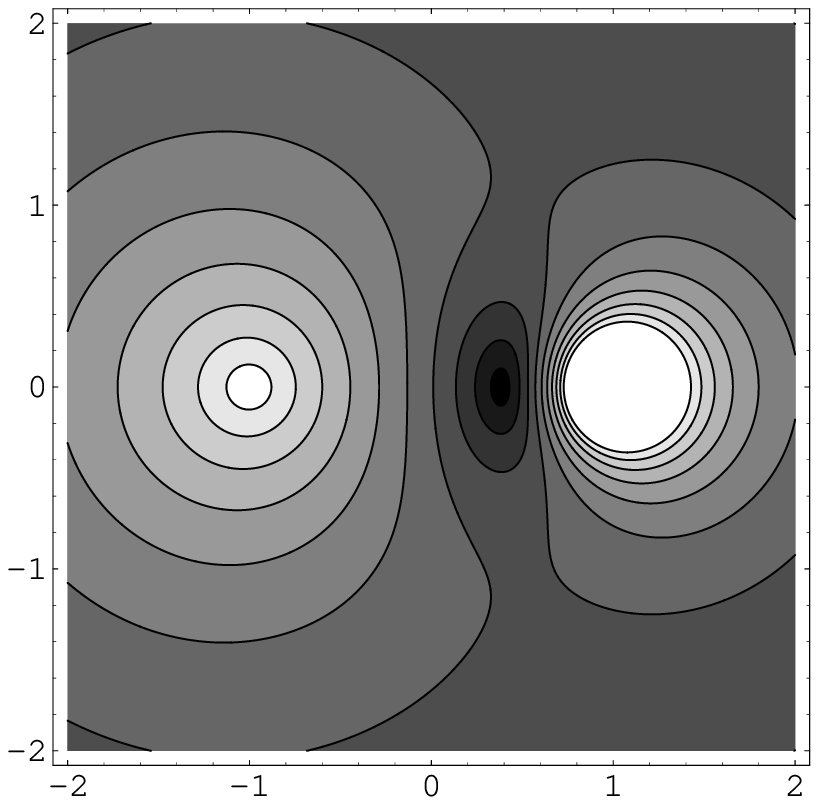},\includegraphics{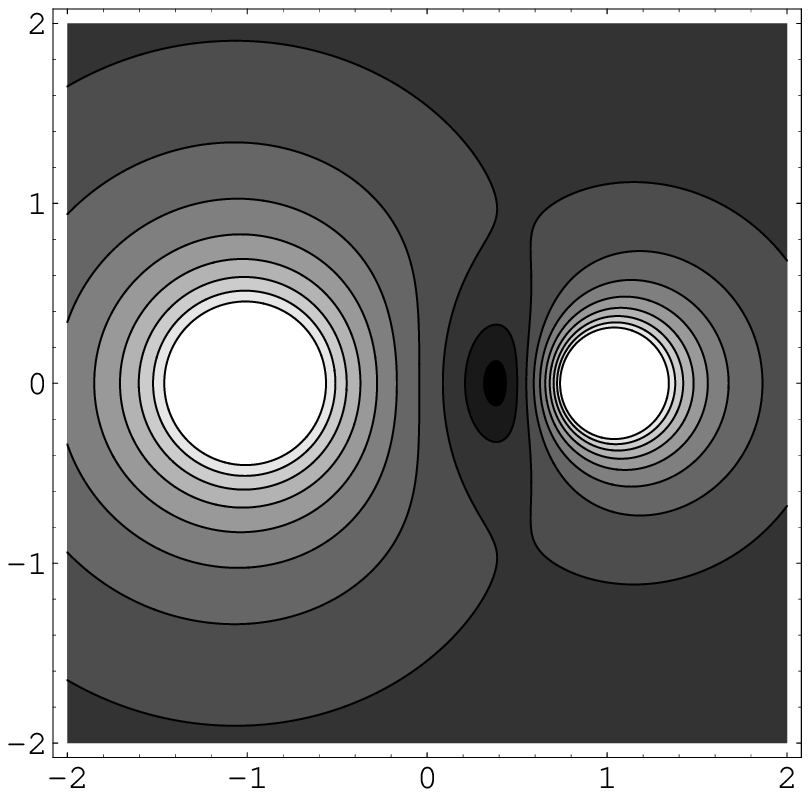}
}
\caption{Temperature for system of two extremely charged black holes with a dilaton in the $(x, y)$-plane: $M_{1} =5/3$, $M_2 =1/3$, $x_1 =-1$, $x_2 =1$;
a) left - $a=0$, b) $a=1$, c) right - $a=3$ ($\hbar=1$).}
\label{CC}
\end{figure*}

Comparison of isotherms for three different values of $a$ ($a=0$ corresponds to the Einstein-Maxwell theory and $a=1$ to string theory) shows that when the dilaton field is "turned on", the space around the heavier black hole is "heated up" (see Fig. \ref{CC}).

\section{Conclusions}

In this work we have considered the acceleration, Unruh temperature and energy on the holographic screens for various static, but not necessarily spherically symmetric, solutions. We have shown the Unruh-Verlinde temperature and acceleration vanish on the throat of a traversable spherically symmetric wormhole, independently on its shape. The holographic equipotential surfaces are shown for less symmetric static solution for a non-rotating black hole immersed in a magnetic field. It has been shown that the Unruh temperature coincides with the Hawking one on the event horizon and that along the axis of the magnetic field there is a region of low temperature.
For a system of two and more extremely charged black holes it has been shown that there is a region of low temperature in the space between the black holes. The presence of a dilaton deform the lines of isotherms of a multi-black hole system, by "heating" the space around the heavier black hole.

In this way we tried to imagine various compact static gravitational objects, such as a black hole with a magnetic field, a wormhole, a system of black holes in equilibrium in this new picture of gravity as an entropic force.
Within Verlinde`s approach to gravity, the analysis of the temperature and equipotential surfaces gains considerable motivation. Therefore we are aimed at rigorous analysis of equipotential surfaces and holographic thermodynamics for space-times with various groups of isometry \cite{inprep}. At the same time portraits of accelerations around gravitating objects, which have been obtained here, might be useful independently on the interpretation of gravity as an entropic force or as a fundamental interaction.

\section*{Acknowledgments}
The author would like to acknowledge support of \emph{the Alexander von Humboldt Foundation}, Germany.
%

\end{document}